\documentstyle [12pt,a4p,epsfig,amsmath,multicol]{article}
\newcommand {\rbf}  {\overline{r}_f}
\newcommand {\vbf}  {\overline{v}_f}
\newcommand {\abf}  {\overline{a}_f}
\newcommand {\sbf}  {\overline{s}_f}

\newcommand {\sbl}  {\overline{s}_l}

\newcommand {\vbb}  {\overline{v}_b}
\newcommand {\abb}  {\overline{a}_b}
\newcommand {\sbb}  {\overline{s}_b}

\newcommand {\gbl}  {\overline{g}_b^L}
\newcommand {\gbr}  {\overline{g}_b^R}

\newcommand {\sleff}  {\sin^2\Theta_{eff}^{lept}}
\begin{document}
\hspace*{8cm} {UGVA-DPNC 1998/10-181 October 1998}
\newline
\hspace*{10.5cm} hep-ph/9810288
\vspace*{0.6cm}
\begin{center}
{\bf The Indirect Limit on the Standard Model Higgs Boson 
 Mass from the Precision FERMILAB, LEP and SLD Data.}
\end{center}
\vspace*{0.6cm}
\centerline{\footnotesize J.H.Field}
\baselineskip=13pt
\centerline{\footnotesize\it D\'{e}partement de Physique Nucl\'{e}aire et 
 Corpusculaire, Universit\'{e} de Gen\`{e}ve}
\baselineskip=12pt
\centerline{\footnotesize\it 24, quai Ernest-Ansermet CH-1211 Gen\`{e}ve 4. }
\centerline{\footnotesize E-mail: john.field@cern.ch}
\baselineskip=13pt
 
\vspace*{0.9cm}
\abstract{ Standard Model fits are performed on the most recent leptonic
 and b quark Z decay data from LEP and SLD, and FERMILAB data on top quark
production, to obtain $m_t$ and $m_H$. Poor fits are obtained, with confidence
levels $\simeq$ 2$\%$. Removing the b quark data improves markedly the
quality of the fits and reduces the 95$\%$ CL upper limit on $m_H$ by 
 $\simeq$ 50 GeV. }
\vspace*{0.9cm}
\normalsize\baselineskip=15pt
\setcounter{footnote}{0}
\renewcommand{\thefootnote}{\alph{footnote}}
\newline 
{\it Keywords ;} Standard Electroweak Model, LEP, SLD and FERMILAB data,
 Z-decays, b quark couplings, top quark and Higgs boson masses
\newline

\vspace*{0.4cm}

\newpage
 
  Since the discovery of the top quark by the CDF and D0 Collaborations
 at FERMILAB
~\cite{x1} and the determination of its mass with a precision of
$\simeq 3\%$~\cite{x2}, an important goal of the analysis of the
 precision electroweak data from LEP and SLD~\cite{x3,x4} has been to
 establish  indirect limits on the mass, $m_H$, of the Standard Model (SM)
 Higgs Boson from the measurement of the effect of quantum corrections
 in Z decays. A 95$\%$ confidence level (CL) lower limit on $m_H$ of
 89.8 GeV has also recently been set in the direct search for the 
 Higgs Boson by the 4 LEP experiments~\cite{x5}. The consistency, or
 otherwise, of the indirect and direct limits for $m_H$ constitutes
 an important test of the SM.  
\par Measurents of the same electroweak observables by different experiments
are combined by the LEP-SLD Electroweak Working Group (LSEWWG)~\cite{x3},
but still, in the global fits to the data used to obtain the indirect
limit on $m_H$, a large number of different `raw' observables are included in
the $\chi^2$. These observables vary widely both in experimental 
 precision and in sensitivity to $m_H$. They may, however, be further 
combined, using only very weak theoretical assumptions (lepton 
universality and the validity of perturbative QED and QCD corrections) to yield
a much smaller number of parameters that contain all precise experimental 
information on $m_H$. Fitting these parameters to the SM prediction,
as is done below, rather than the raw observables, as in the LSEWWG
fits, results in  much sharper test and, as will be seen, clearly
pin-points possible anomalies or inconstencies in the data. There are
 essentially
four such independent parameters, which may be chosen to be the effective 
weak coupling constants (vector and axial vector, or right-handed and
left-handed) of the charged leptons and b quarks. The effective coupling
constants of the other quarks have a similar theoretical status but, 
because of their much larger experimental errors, have a negligible 
weight in the determination of $m_H$
\footnote{Although the direct measurement of the W mass is 
 expected, in the future, to provide valuable information on
 $m_H$, the present experimental error is too large to be 
 competitive with Z decay measurements.}. Actually, in the SM, although all 
four parameters are sensitive to $m_t$ given the present experimental
errors, the sensitivity of the b quark couplings to $m_H$ is extremely
 weak.
 The method of extraction of the
effective coupling constants from the raw observables as been described
previously~\cite{x6,x7,x8}. In order to simplify the fitting procedure
it is convenient to use, instead of the effective vector (axial vector)
coupling constants $\vbf$ ($\abf$) ($f=l,b$) the equivalent quantities,
with uncorrelated experimental errors, $A_f$, $\sbf$ defined by the
relations:
 \begin{equation}
    A_f \equiv \frac{2 (\sqrt{1-4 \mu_f}) \rbf}{1-4 \mu_f+(1+2 \mu_f) \rbf^2},
    \end{equation} 
 where 
\[  \rbf \equiv \vbf/\abf, \]
 and 
 \begin{equation}
 \overline{s}_f \equiv (\overline{a}_f)^2(1-6 \mu_f)+(\overline{v}_f)^2.
 \end{equation} 
 The parameter  $\mu_f = (\overline{m}_f(M_Z)/M_Z)^2$ where
$\overline{m}_f(Q)$ is the running fermion mass at the scale $Q$, can be set
to zero for $f=l$ to sufficient accuracy, while for b quarks 
$(\overline{m}_b(M_Z)/M_Z)^2 = 1.0 \times 10^{-3}$~\cite{x9}. The values of
$A_l$, $\sbl$, $A_b$, $\sbb$ extracted from the most recent compilation of
electroweak data~\cite{x4} are presented in Table 1 where they are compared
with the SM prediction for $m_t = 174$ GeV, $m_H = 100$ GeV. The SM
 predictions used here are derived from the ZFITTER5.10 program
 package~\cite{x10}, which includes the recently calculated
$O(g^4m_t^2/M_W^2)$ two-loop corrections~\cite{x11}. Good agreement is
 seen for all parameters except $A_b$, which differs from the SM prediction
 by 3.0 standard deviations. The CL that all four parameters agree with
 the SM is only 1.0$\%$ ($\chi^2/dof = 13.2/4$). This apparent
 anomaly was already apparent in the 1996 LSEWWG averages~\cite{x12}, and has
 been extensively discussed~\cite{x6,x7}.The right-handed (R) and
 left-handed (L) 
effective couplings of the b quarks:
~$\gbr = (\vbb-\abb)/2$, $\gbl = (\vbb+\abb)/2$
are found to have the values:
\[ \gbr = 0.1050(90),~~~~~~\gbl = -0.4159(24) \]
as compared with the respective SM predictions of 0.0774 and -0.4208.  
The largest anomaly is in $\gbr$ (3.1$\sigma$) rather than $\gbl$ (2.0$\sigma$).
\par The purpose of this letter is twofold: (i) To recall that only one parameter,
$\gbr$, among the four that contain all the high precision information on 
quantum corrections in Z decays shows a large deviation from the
 SM prediction~\cite{x6}.
(ii) To point out that the values of the limits on $m_H$ depend strongly on 
inclusion or exclusion of the b quark data. Using only the leptonic data, 
that agrees well with the SM prediction, leads to significantly lower values
of $m_H$.
\par The results of SM fits for $m_H$ and $m_t$ to the parameter sets
$A_l$, $\sbl$, $m_t$ and $A_l$, $\sbl$, $A_b$, $\sbb$, $m_t$ are presented in 
 Table 2. The recent CDF, D0 average~\cite{x2,x4}: $m_t = 173.8 \pm 5.0$ GeV and the
 fixed value $\alpha_s(M_Z) = 0.120$, consistent with
 the world average $0.118(5)$~\cite{x13,x14} 
 is used in the fits. For each parameter set three fits are performed for
 different values of $\alpha(M_Z)$, corresponding to the experimental value:
 $\alpha(M_Z)^{-1}=128.896(90)$~\cite{x15}, and $\pm$ 1$\sigma$ variations on
the value. The fitted value of $m_H$ is seen to be very sensitive to 
$\alpha(M_Z)$. All fits give a very stable value of $m_t$ of $\simeq 171.2$ with
 a maximum variation of 0.7 GeV, much smaller than the typical fit error of
$\simeq 3.7$ GeV. On the other hand, large variations are seen in $m_H$ both as
a function of $\alpha(M_Z)$ and on the inclusion or exclusion of the b quark data.
For $\alpha(M_Z)^{-1}=128.896$ the fit excluding the b quark data gives 
$m_H=38.0_{-19.8}^{+30.5}$ and a 95$\%$ CL upper limit of $94$ GeV; including
the b quark data gives $m_H = 77.8_{-26.2}^{+38.6}$ and an upper limit of $150$ GeV.
The CLs of the SM fits to the lepton data and $m_t$ are in the range $24\%-57\%$, 
whereas when the b quark data is included, the CLs drop to only $1.7\%-1.8\%$.
 The results
on the indirect Higgs Boson Masss limits are summarised Table 3, where the
variations due to the experimental error on $\alpha(M_Z)$ and $\pm$ 1$\sigma$
variations in the fitted value of $m_t$ are also presented.
 When the b quark data is included, the `maximum'\footnote{Given by adding
 linearly the shifts generated by the experimental error on $\alpha(M_Z)$
 and the fit error on $m_t$.} 95$\%$ CL upper limit on $m_H$ is found to 
 be 278 GeV, in good agreement with the LSEWWG value of 280 GeV~\cite{x4}.
 Excluding the b quark
data, which is incompatible, at the 3$\sigma$ level, with the SM, reduces the
fitted value of $m_H$ by a factor two, and lowers the 95$\%$ CL upper limit by
$56$ GeV. Taking into acccount the
 strong dependence of the limit on $\alpha(M_Z)$ and
$m_t$ (see Table 3),
 this is still quite consistent with the direct lower limit of $89.9$ GeV
~\cite{x5}. It should be stressed that the shift in the value of $m_H$ is generated
 due to the high sensitivity of $A_l$ via correlations (
 $A_{FB}^{0,b}=3 A_l A_b / 4$) and not by any variation
 in the quantity $A_b$, which is quite insensitive to $m_H$.
\begin{figure}[htbp]
\begin{center}\hspace*{-0.5cm}\mbox{
\epsfysize10.0cm\epsffile{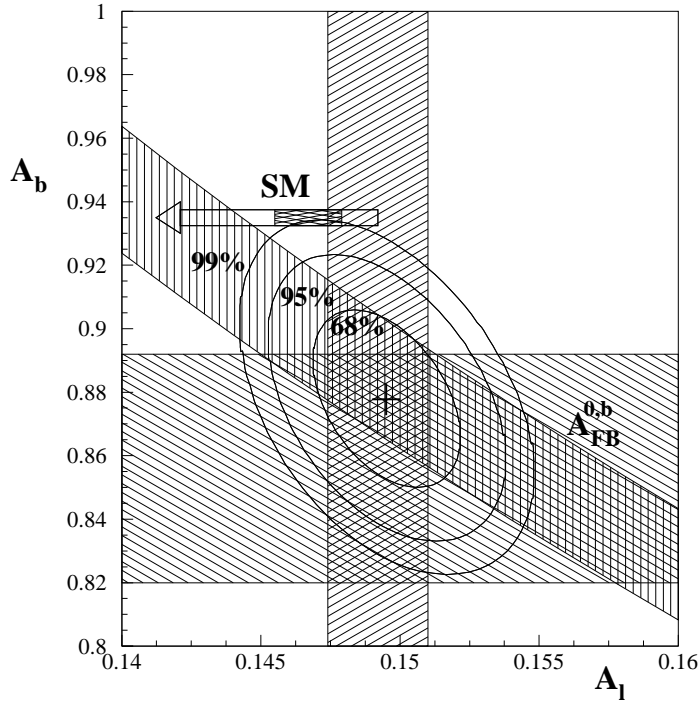}}
\caption{ The cross-hatched bands show the $\pm 1 \sigma$ limits for the
 quantities $A_l(\rm{LEP+SLD}) $, $A_b(\rm{SLD})$. and $A_{FB}^{0,b}(\rm{LEP})$.
 The cross shows the best fit to $A_l$ and $A_b$, togther with 68$\%$, 95$\%$
 and 99$\%$ CL contours. The narrow cross hatched rectangle shows the SM
 prediction for $m_H = 100$ GeV and $m_t = 174 \pm 5$ GeV. The open
 arrow shows the SM prediction for $m_H = 100_{-50}^{+200}$ GeV and $m_t = 174$ GeV.
 The arrow points in the direction of increasing $m_H$. }
\label{fig-fig1}
\end{center}
 \end{figure}  
 This point is made clear by Figure 1, which shows a two dimensional plot
 of the LEP+SLD average value $A_l$ and $A_b(\rm{SLD})$. The diagonal band
 shows the LEP $A_{FB}^{0,b}$ measurement. Also shown are the 68$\%$, 95$\%$
 and 99$\%$ CL contours of the best fit to $A_l$ and $A_b$ using all three
 data, as well as the prediction of the SM that lies just
 outside the 99$\%$ CL contour. The shift towards higher values
  of $m_H$ caused by
 the  $A_{FB}^{0,b}$ measurement as well as poor agreement of the fit 
 with the SM are evident.  
\par None of the above conclusions were reported when the results of global SM 
fits by the LSEWWG to the same data set used in this letter, were presented at
the recent Vancouver conference~\cite{x4}. This is because no attempt was made 
to extract the effective couplings of the b quarks, and the SM fit was performed
on a large number (20) of raw electroweak observables, many of which have large
errors and/or are relatively insensitive to $m_H$ or the b quark couplings.
 In fact it is clear from inspection of Figure 1 that the 3 largest `pulls'
 \footnote{ i.e. (measurement-fit)/error} in the global EW fit shown in
 Ref.[4](due to $A_b(\rm{SLD})$, $A_{FB}^{0,b}$ and $\sleff$ derived from $A_{LR}$), are
 all correlated to the large deviation of the best fit value of $A_b$ from
 the SM prediction. These three data alone contribute 11.1 (or 65$\%$) out of
 the total $\chi^2$ of 17.0 for 15 $dof$. 
  The 3.1$\sigma$ deviation of $\gbr$ from the SM is not revealed
 in the SLEWWG fit. Instead smaller deviations appear in the correlated
 quantities $A_b(\rm{SLD})$, $A_{FB}^{0,b}$ and $R_b$. 
 It is interesting to note that
 the 17 data whose pulls are least effected by the deviation in the b
 quark couplings
 give an anomalously low contribution to the $\chi^2$
 ($\chi^2/dof = 5.9/17$, CL=99.45$\%$) indicating that, on average, the errors for
 these quantities may be overestimated by a factor of $\simeq 1.7$.
The very low contribution from these data hides the large
positive contribution resulting from the deviation in $A_b$ when only
the global $\chi^2$ is considered. 
 A similar criticism may be made of another
 recent global analysis~\cite{x16} based on the data set used in this letter.
 In this case the global $\chi^2$ contained 42 data fit to 6 parameters (including
$m_t$ and $m_H$) yielding a $\chi^2/dof = 28.8/36$ (CL $= 80\%$ ). It is
stated, in consequence, that: `The fit to all precision data is perfect'.
 Although it is true that, as in the SLEWWG fit, `None of the observables
deviates from the SM best fit prediction by more than 2 standard deviations'
it also remains true that an anomalously large contribution to the $\chi^2$
 comes from the b quark data, where the effective couplings 
{\it do deviate from the SM at the 3$\sigma$ level}. This is completely hidden
by the good ageement with the SM of 39 out of the 42 data that are fitted!
\par Finally, it may be mentioned that none of the previous discussions
in the literature of the sensitivity of $m_H$ to different data sets
~\cite{x17,x18,x19} pointed out either the sensitivity of the limit to
the b quark data, or the poor overall confidence levels of SM fits to the
effective couplings when the latter are included. A more detailed
discussion of this previous literature is given in Ref.[8].
\newline
{\bf Acknowledgements}
\newline 
I thank M.Dittmar for discussions, and his encouragement for the pursuit of
this work.

\pagebreak

\pagebreak

\begin{table}
\begin{center}
\begin{tabular}{|c|c|c|c|c|} \cline{2-5}
\multicolumn{1}{c}{ } & \multicolumn{2}{|c}{leptons }
 & \multicolumn{2}{|c|}{ b quarks }  \\ \cline{2-5}
\multicolumn{1}{c|}{ } & $A_l$  & $\sbl$  &  $A_b$ & $\sbb$  \\ \hline
 Meas. & 0.1492(18) & 0.25243(30)  &
0.878(19)  & 0.3662(14) \\ \hline
 SM &  0.1467 & 0.25272  & 0.9347 & 0.3647 \\ \hline
Dev.($\sigma$) & 1.4 & -1.0  & -3.0 & 1.1 \\
 \hline      
\end{tabular}
\caption[]{ Measured values of $A_f$ and $\sbf$ ($f=l,b$) compared to SM 
 predictions for $m_t =$ 174 GeV, $m_H =$ 100 GeV.  Dev($\sigma$) = (Meas.-SM)/Error. } 
\end{center}
\end{table}

\begin{table}
\begin{center}
\begin{tabular}{|c|c|c|c|c|} \hline
Fitted Quantities & $\alpha(M_Z)^{-1}$  & $m_t$ (GeV) & $m_H$ (GeV) & C.L.($\%$)  \\ \hline
\multicolumn{1}{|c|}{ } & 128.986 & $171.5 \pm 3.8$ &
 $73.8_{-29.4}^{+46.5}$ [166] & 24  \\ \cline{2-5}
$A_l$, $\sbl$, $m_t$ & 128.896  & $170.7 \pm 3.8$ &
$38.0_{-19.8}^{+30.5}$ [94] & 28  \\ \cline{2-5}
\multicolumn{1}{|c|}{ } & 128.806 &  $172.0 \pm 3.8$ &
 $19.6_{-8.0}^{+18.1}$ [54] & 57 \\ \hline
\multicolumn{1}{|c|}{ } & 128.986 & $171.9 \pm 3.6$ &
 $124.7_{-40.9}^{+58.7}$ [234] & 1.8  \\ \cline{2-5}
 $A_l$, $\sbl$, $A_b$, $\sbb$, $m_t$ & 128.896 &
 $171.4 \pm 3.6$ & $77.8_{-26.2}^{+38.6}$ [150] & 1.7 \\ \cline{2-5}
\multicolumn{1}{|c|}{ } & 128.806 &  
 $171.3 \pm 3.6$ & $44.1_{-19.6}^{+22.5}$ [87] & 1.8 \\ \hline 
\end{tabular}
\caption[]{ SM fits to different data sets. 95$\%$ C.L. upper limits for $m_H$ are 
given in the square brackets. } 
\end{center}
\end{table}
\begin{table}
\begin{center}
\begin{tabular}{|c|c|c|} \hline
Fitted Quantities & $m_H$ (GeV)  & 95 $\%$ CL upper limit on $m_H$ (GeV)   \\ \hline
$A_l$, $\sbl$, $m_t$ & $38_{-20 -18 -9.5}^{+31 +36 +17}$  &  $94_{-40 -23}^{+72 +34}$  \\ \hline 
 $A_l$, $\sbl$, $A_b$, $\sbb$, $m_t$ & $78_{-26 -34 -17}^{+39 +47 +24}$ & $150_{-63 -33}^{+84 +44}$
 \\ \hline
\end{tabular}
\caption[]{ Summary of SM fit results for $m_H$. The errors on $m_H$ are, in order:
 the 1$\sigma$ fit error, and the changes produced by 
 $\pm 1\sigma$ variations in $\alpha(M_Z)^{-1}$
 and $m_t$. The errors on the upper limit are those due to $\pm 1\sigma$ variations
 in $\alpha(M_Z)^{-1}$ and $m_t$.}
\end{center}
\end{table}

\begin{thebibliography}{99}
\bibitem{x1}
F.Abe et al. The CDF Collaboration, Phys. Rev. {\bf D56} 5919 (1974),
Phys. Rev. Lett. {\bf 74} 2626 (1995). S.Abachi et al. The D0 Collaboration,
Phys. Rev. Lett. {\bf 74} 2632 (1995).
\bibitem{x2}
Review of Particle Properties, Particle Data Group,
C.Caso et al. Eur. Phys. J.  {\bf C3} 1 (1998).
\bibitem{x3}
The LEP Collaborations ALEPH, DELPHI, L3, OPAL,
the LEP Electroweak Working Group and the SLD
Heavy Flavour Group. LEPEWWG/97-01 (1997).
\bibitem{x4}
M.Gr\"{u}newald and D.Karlen, in proceedings of the XXIX International
 Conference on High Energy Physics, UBC, Vancouver, BC, Canada, 
July 23-29 1998.  
\bibitem{x5}
ALEPH, DELPHI, L3 and OPAL Collaborations, 
`Lower bound on the Standard Model Higgs boson mass 
 from combining the results of the four LEP experiments',
 CERN-EP/98-046.   
\bibitem{x6}
J.H.Field, Mod. Phys. Lett. A, Vol. 13, No. 24, 1937  (1998).
\bibitem{x7}
J.H.Field, Phys. Rev. {\bf D58} 093010-1 (1998).
\bibitem{x8}
J.H.Field {\it `Z-decays to b Quarks and the Higgs Boson Mass'} UGVA-DPNC 1998/09-179
 September 1998 (hep-ph/9809292). Submitted to Phys. Rev. D.
\bibitem{x9}
G.Rodrigo, Nucl. Phys. B (Proc. Suppl.) {\bf 54A}, 60 (1997).
\bibitem{x10}
D.Bardin et al. FORTRAN package ZFITTER, Preprint CERN-TH 6443/92.
\bibitem{x11}
G.Degrassi, P.Gambino  and A.Vicini, Phys. Lett. {\bf B383}, 219 (1996),
G.Degrassi, P.Gambino  and A.Sirlin, Phys. Lett. {\bf B394}, 188 (1997),
G.Degrassi, et al. Phys. Lett. {\bf B418}, 209 (1998).
\bibitem{x12}
The LEP-SLD Electroweak Working Group (see Ref.[3]),
 CERN-PPE/96-183 (1996).
\bibitem{x13}
M.Schmelling in Proceedings of the 28th International
Conference on High Energy Physics, 
Warsaw 1996, Eds. Z.Ajduk and A.K.Wroblewski.
\bibitem{x14}
P.N.Burrows, Talk presented at the 3rd International Symposium 
on Radiative Corrections, August 1-5 1996, Cracow, Poland.
Pre-print: SLAC-PUB-7293.
\bibitem{x15}
S.Eidelmann and F.Jegerlehner, Z. Phys. {\bf C67}, 585 (1995).
\bibitem{x16}
J.Erler and P.Langacker {\it `Status of the Standard Model'} Pre-print UPR-0816-T,
(hep-ph/9809352).
\bibitem{x17}
A.Gurtu, Phys. Lett. {\bf B368}, 247 (1996).
\bibitem{x18}
S.Dittmaier, D.Schildknecht and G.Weiglen, {\bf B385}, 415 (1996).
\bibitem{x19}
M.S.Chanowitz, Phys. Rev. Letters {\bf 80}, 2521 (1998),
and pre-print LBNL-42103 (hep-ph/9807452). 
\end{thebibliography}
\end{document}